\documentclass[aps,prd]{revtex4}
\usepackage{latexsym}
\usepackage{graphicx}

\newcommand{\cc}{cosmological constant}
\newcommand{\del}{\partial}
\newcommand{\dphi}{\partial_i \phi \partial^i \phi}

\begin{document}

\title{ Dark Energy: the Cosmological Challenge of the Millennium}

\author{T.~Padmanabhan}
\affiliation{IUCAA, 
Post Bag 4, Ganeshkhind, Pune - 411 007\\
email: nabhan@iucaa.ernet.in}

\begin{abstract}

Recent cosmological observations suggest that nearly seventy  per cent of the energy density in the universe is unclustered and  has negative pressure. Several conceptual issues related to the modeling of this component (`dark energy'), which is driving an accelerated expansion of the universe, are discussed with special emphasis on the cosmological constant as the possible choice for the dark energy. Some curious geometrical features of a universe with a cosmological constant are
described and a few attempts to understand the nature of the cosmological constant are reviewed. 

 \end{abstract}

\maketitle

\section{The Cosmological Paradigm}

The last couple of decades have been the golden age for cosmology, in much the same way as the mid-1900's
were a golden age for particle physics. Data of exquisite quality confirmed the broad paradigm of standard
 cosmology and helped us to determine the composition of  the universe. As a direct consequence, the cosmological observations have thrusted upon us a rather preposterous
composition for the universe which defies any simple explanation, thereby posing the greatest challenge
theoretical physics has ever faced.

To understand these exciting developments, it is best to begin by reminding ourselves of the standard
paradigm for cosmology. Observations show that the universe is fairly homogeneous and isotropic at scales
larger than about $150h^{-1}$ Mpc,  where 1 Mpc$\approx 3\times 10^{24}$ cm is a convenient unit for extragalactic astronomy and $h\approx 0.7$ characterizes  \cite{h} the current rate of expansion of the universe in dimensionless form. (The mean distance between galaxies is about 1 Mpc while the size of the visible universe is about $3000 h^{-1}$ Mpc.) The conventional --- and highly successful --- approach to cosmology
separates the study of large scale ($l\gtrsim 150h^{-1}$ Mpc) dynamics of the universe from the issue of structure formation at smaller scales. The former is modeled by  a homogeneous and isotropic distribution of energy density; the latter issue is
addressed  in terms of gravitational instability which will amplify the small perturbations in the energy density, leading to the formation of structures like galaxies.

In such an approach, the expansion of the background universe is described by a single function of time $a(t)$
which is governed by the equations (with $c=1$): 
\begin{equation} 
\frac{\dot a^2+k}{a^2} =\frac{8\pi G\rho}{3};\qquad d(\rho a^3)=-pda^3
\label{frw}
\end{equation}
The first one relates expansion rate to the energy density $\rho$ and $k=0,\pm 1$ is a parameter which characterizes the spatial curvature of the universe. 
 The second equation, when coupled with the equation of state
$p=p(\rho)$, determines the evolution of energy density  $\rho=\rho(a)$ in terms of the expansion factor of the universe.
 In particular if $p=w\rho$ with (at least, approximately) constant $w$ then, $ \rho \propto a^{-3(1+w)}$ and (if we  assume $k=0$), 
$ a \propto t^{2/[3(1+w)]}$.
 
 It is convenient to measure
the energy densities of different components in terms of a \textit{critical energy density} ($\rho_c$) required to make $k=0$ at the present epoch. (Of course, since $k$ is a constant,
it will remain zero at all epochs if it is zero at any given moment of time.) From Eq.(\ref{frw}), it is clear that $\rho_c=3H^2_0/8\pi G$ where $H_0=(\dot a/a)_0$
is the rate of expansion of the universe at present.  The variables $\Omega_i=\rho_i/\rho_c$ 
will give the fractional contribution of different components of the universe ($i$ denoting baryons, dark matter, radiation, etc.) to the  critical density. Observations then lead to the following results:

\begin{itemize}
\item
Our universe has $0.98\lesssim\Omega_{tot}\lesssim1.08$. The value of $\Omega_{tot}$ can be determined from the angular anisotropy spectrum of the cosmic microwave background radiation (CMBR) (with the reasonable assumption that $h>0.5$) and these observations now show that we live in a universe
with critical density \cite{cmbr,kanduhere}. 
\item
Observations of primordial deuterium produced in big bang nucleosynthesis (which took place when the universe
was about 1 minute in age) as well as the CMBR observations show that  \cite{baryon} the {\it total} amount of baryons in the
universe contributes about $\Omega_B=(0.024\pm 0.0012)h^{-2}$. Given the independent observations on the Hubble constant \cite{h} which fix $h=0.72\pm 0.07$, we conclude that   $\Omega_B\cong 0.04-0.06$. These observations take into account all baryons which exist in the universe today irrespective of whether they are luminous or not. Combined with previous item we conclude that
most of the universe is non-baryonic.
\item
Host of observations related to large scale structure and dynamics (rotation curves of galaxies, estimate of cluster masses, gravitational lensing, galaxy surveys ..) all suggest \cite {dm} that the universe is populated by a non-luminous component of matter (dark matter; DM hereafter) made of weakly interacting massive particles which \textit{does} cluster at galactic scales. This component contributes about $\Omega_{DM}\cong 0.20-0.35$.
\item
Combining the last observation with the first we conclude that there must be (at least) one more component 
to the energy density of the universe contributing about 70\% of critical density. Early analysis of several observations
\cite{earlyde} indicated that this component is unclustered and has negative pressure. This is confirmed dramatically by the supernova observations (see \cite{sn}; for a critical look at the data, see \cite{tptirthsn1,tptirthsn2}).  The observations suggest that the missing component has 
$w=p/\rho\lesssim-0.78$
and contributes $\Omega_{DE}\cong 0.60-0.75$.
\item
The universe also contains radiation contributing an energy density $\Omega_Rh^2=2.56\times 10^{-5}$ today most of which is due to
photons in the CMBR. This is dynamically irrelevant today but would have been the dominant component in the universe  at redshifts
larger that $z_{eq}\simeq \Omega_{DM}/\Omega_R\simeq 4\times 10^4\Omega_{DM}h^2$.
\item
Together we conclude that our universe has (approximately) $\Omega_{DE}\simeq 0.7,\Omega_{DM}\simeq 0.26,\Omega_B\simeq 0.04,\Omega_R\simeq 5\times 10^{-5}$. All known observations
are consistent with such an --- admittedly weird --- composition for the universe.
\end{itemize}

Before discussing the composition of the universe in greater detail, let us briefly consider the issue of structure formation. The key idea is that if there existed small fluctuations in the energy density in the early universe, then gravitational instability can amplify them in a well-understood manner (see e.g., \cite{tpsfuv3}), leading to structures like galaxies etc. today. The most popular theoretical model for these fluctuations is based on the idea that if the very early universe went through an inflationary phase \cite{inflation}, then the quantum fluctuations of the field driving the inflation can lead to energy density fluctuations\cite{genofpert,tplp}. It is possible to construct models of inflation such that these fluctuations are described by a Gaussian random field and are characterized by a power spectrum of the form $P(k)=A k^n$ with $n\simeq 1$. The models cannot predict the value of the amplitude $A$ in an unambiguous manner but it can be determined from CMBR observations. The CMBR observations are consistent with the inflationary model for the generation of perturbations and gives $A\simeq (28.3 h^{-1} Mpc)^4$ and $n=0.97\pm0.023$ (The first results were from COBE \cite{cobeanaly} and
WMAP  \cite{kanduhere} has reconfirmed them with far greater accuracy).

So, to the zeroth order, the universe is characterized by just seven numbers: $h\approx 0.7$ describing the current rate of expansion; $\Omega_{DE}\simeq 0.7,\Omega_{DM}\simeq 0.26,\Omega_B\simeq 0.04,\Omega_R\simeq 5\times 10^{-5}$ giving the composition of the universe; the amplitude $A\simeq (28.3 h^{-1} Mpc)^4$ and the index $n\simeq 1$ of the initial perturbations. 
The challenge is to make some sense out of these numbers from a more fundamental point of view.

\section{The Dark Energy}

It is rather frustrating that we have no direct laboratory evidence for nearly 96\% of matter in the universe.
(Actually, since we do not quite understand the process of baryogenesis, we do not understand $\Omega_B$ either;
all we can \textit{theoretically} understand now is a universe filled entirely with radiation!). Assuming that particle physics models will eventually come of age and (i) explain  $\Omega_B$ and $\Omega_{DM}$ (probably as the lightest
supersymmetric partner) as well as (ii) provide a viable model for inflation predicting correct value for $A$,
one is left with the problem of understanding $\Omega_{DE}$. While the issues (i) and (ii) are by no means trivial or satisfactorily addressed, it is probably correct to say that the issue of dark energy is lot more perplexing,
thereby justifying the attention it has received recently.

\begin{figure}[ht]
\begin{center}
\includegraphics[angle=-90,scale=.5]{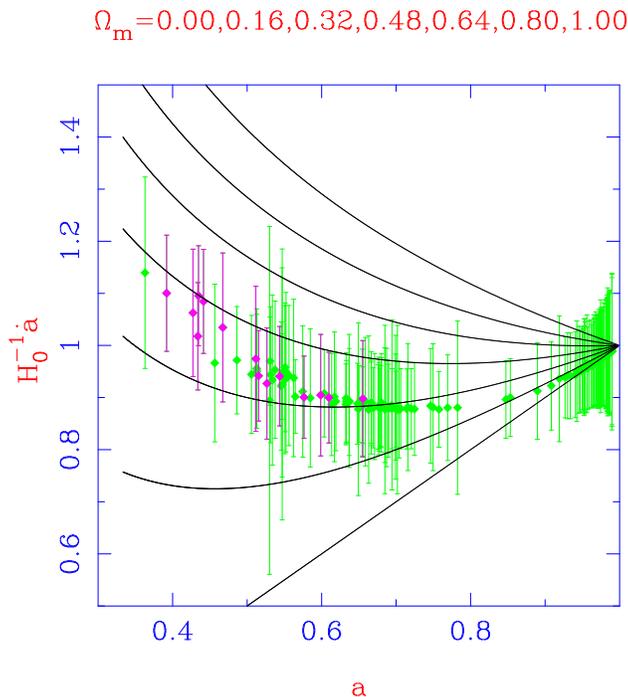}
\end{center}
\caption{The ``velocity'' $\dot a$  of the universe is plotted against the ``position'' $a$ in the form of a phase portrait. The different curves are for models parameterized by the value of  $\Omega_{DM}(=\Omega_m)$ keeping $\Omega_{tot}=1$. The top-most curve has $\Omega_m=1$ with no dark energy and the universe is decelerating at all epochs. The bottom-most curve has $\Omega_m=0$
and $\Omega_{DE}=1$ and the universe is accelerating at all epochs. The in-between curves show  universes which were decelerating in the past and began to accelerate when the dark energy started dominating. The supernova data clearly favours such a model. (For a more detailed discussion of the figure, see
\cite{tptirthsn1,tptirthsn2}.)}
\label{fig:tptrc}
\end{figure}

The key observational feature of dark energy is that --- treated as a fluid with a stress tensor $T^a_b={\rm dia} (\rho, -p, -p,-p)$ 
--- it has an equation state $p=w\rho$ with $w \lesssim -0.8$ at the present epoch. 
The spatial part  ${\bf g}$  of the geodesic acceleration (which measures the 
  relative acceleration of two geodesics in the spacetime) satisfies an \textit{exact} equation
  in general relativity (see e.g., page 332 of \cite{probbook}) given by:
  \begin{equation}
  \nabla \cdot {\bf g} = - 4\pi G (\rho + 3p)
  \label{nextnine}
  \end{equation} 
 This  shows that the source of geodesic  acceleration is $(\rho + 3p)$ and not $\rho$.
  As long as $(\rho + 3p) > 0$, gravity remains attractive while $(\rho + 3p) <0$ can
  lead to repulsive gravitational effects. In other words, dark energy with sufficiently negative pressure will
  accelerate the expansion of the universe, once it starts dominating over the normal matter.  This is precisely what is established from the study of high redshift supernova, which can be used to determine the expansion
rate of the universe in the past \cite{sn}. 
Figure \ref{fig:tptrc} presents the supernova data as a phase portrait \cite{tptirthsn1,tptirthsn2} of the universe (plotting the `velocity' $\dot a$ against 'position' $a$). It is  clear that the universe was decelerating at high redshifts and started accelerating when it was about two-third of the present size. 

The simplest model for  a fluid with negative pressure is the
cosmological constant (for a review, see \cite{cc}) with $w=-1,\rho =-p=$ constant (which is the model used in Figure \ref{fig:tptrc}).
If the dark energy is indeed a cosmological constant, then it introduces a fundamental length scale in the theory $L_\Lambda\equiv H_\Lambda^{-1}$, related to the constant dark energy density $\rho_{DE}$ by 
$H_\Lambda^2\equiv (8\pi G\rho_{DE}/3)$.
In classical general relativity,
    based on the constants $G, c $ and $L_\Lambda$,  it
  is not possible to construct any dimensionless combination from these constants. But when one introduces the Planck constant, $\hbar$, it is  possible
  to form the dimensionless combination $H^2_\Lambda(G\hbar/c^3) \equiv  (L_P^2/L_\Lambda^2)$.
  Observations then require $(L_P^2/L_\Lambda^2) \lesssim 10^{-123}$.
  As has been mentioned several times in literature, this will require enormous fine tuning. What is more,
 in the past, the energy density of 
  normal matter and radiation  would have been higher while the energy density contributed by the  cosmological constant
  does not change.  Hence we need to adjust the energy densities
  of normal matter and cosmological constant in the early epoch very carefully so that
  $\rho_\Lambda\gtrsim \rho_{\rm NR}$ around the current epoch.
  This raises the second of the two cosmological constant problems:
  Why is it that $(\rho_\Lambda/ \rho_{\rm NR}) = \mathcal{O} (1)$ at the 
  {\it current} phase of the universe ?

  Because of these conceptual problems associated with the cosmological constant, people have explored a large variety of alternative possibilities. The most popular among them uses a scalar field $\phi$ with a suitably chosen potential $V(\phi)$ so as to make the vacuum energy vary with time. The hope then is that, one can find a model in which the current value can be explained naturally without any fine tuning.
  A simple form of the source with variable $w$ are   scalar fields with
  Lagrangians of different forms, of which we will discuss two possibilities:
    \begin{equation}
  L_{\rm quin} = \frac{1}{2} \partial_a \phi \partial^a \phi - V(\phi); \quad L_{\rm tach}
  = -V(\phi) [1-\partial_a\phi\partial^a\phi]^{1/2}
  \label{lquineq}
  \end{equation}
  Both these Lagrangians involve one arbitrary function $V(\phi)$. The first one,
  $L_{\rm quin}$,  which is a natural generalization of the Lagrangian for
  a non-relativistic particle, $L=(1/2)\dot q^2 -V(q)$, is usually called quintessence (for
  a sample of models, see \cite{phiindustry}).
    When it acts as a source in Friedman universe,
   it is characterized by a time dependent
  $w(t)$ with
    \begin{equation}
  \rho_q(t) = \frac{1}{2} \dot\phi^2 + V; \quad p_q(t) = \frac{1}{2} \dot\phi^2 - V; \quad w_q
  = \frac{1-(2V/\dot\phi^2)}{1+ (2V/\dot\phi^2)}
  \label{quintdetail}
  \end{equation}

The structure of the second Lagrangian  in Eq.~(\ref{lquineq}) can be understood by a simple analogy from
special relativity (see the first reference in \cite{tptirth}). A relativistic particle with  (one dimensional) position
$q(t)$ and mass $m$ is described by the Lagrangian $L = -m \sqrt{1-\dot q^2}$.
It has the energy $E = m/  \sqrt{1-\dot q^2}$ and momentum $k = m \dot
q/\sqrt{1-\dot q^2} $ which are related by $E^2 = k^2 + m^2$.  As is well
known, this allows the possibility of having \textit{massless} particles with finite
energy for which $E^2=k^2$. This is achieved by taking the limit of $m \to 0$
and $\dot q \to 1$, while keeping the ratio in $E = m/  \sqrt{1-\dot q^2}$
finite.  The momentum acquires a life of its own,  unconnected with the
velocity  $\dot q$, and the energy is expressed in terms of the  momentum
(rather than in terms of $\dot q$)  in the Hamiltonian formulation. We can now
construct a field theory by upgrading $q(t)$ to a field $\phi$. Relativistic
invariance now  requires $\phi $ to depend on both space and time [$\phi =
\phi(t, {\bf x})$] and $\dot q^2$ to be replaced by $\partial_i \phi \partial^i
\phi$. It is also possible now to treat the mass parameter $m$ as a function of
$\phi$, say, $V(\phi)$ thereby obtaining a field theoretic Lagrangian $L =-
V(\phi) \sqrt{1 - \del^i \phi \del_i \phi}$. The Hamiltonian  structure of this
theory is algebraically very similar to the special  relativistic example  we
started with. In particular, the theory allows solutions in which $V\to 0$,
$\dphi \to 1$ simultaneously, keeping the energy (density) finite.  Such
solutions will have finite momentum density (analogous to a massless particle
with finite  momentum $k$) and energy density. Since the solutions can now
depend on both space and time (unlike the special relativistic example in which
$q$ depended only on time), the momentum density can be an arbitrary function
of the spatial coordinate. This provides a rich gamut of possibilities in the
context of cosmology.
 \cite{tptachyon,tptirth,bjp,tachyon},
  This form of scalar field arises  in string theories \cite{asen} and --- for technical reasons ---
   is called a tachyonic scalar field.
   (The structure of this Lagrangian is similar to those analyzed in a wide class of models
   called {\it K-essence}; see for example, \cite{kessence})

   The stress tensor for the tachyonic scalar  field can be written 
as  the sum of a pressure less dust component and a cosmological constant (see the first reference in \cite{tptirth}).
To show this explicitly,
we  break up the density $\rho$ and the pressure $p$
and write them in a more suggestive form as
$\rho = \rho_\Lambda  + \rho_{\rm DM}; ~~
p = p_V  + p_{\rm DM}$
where
\begin{equation}
\rho_{\rm DM} = \frac{V(\phi) \del^i \phi \del_i \phi}
{\sqrt{1 - \del^i \phi \del_i \phi}};\ \  p_{\rm DM} = 0;\ \
\rho_\Lambda  = V(\phi) \sqrt{1 - \del^i \phi \del_i \phi};\ \
p_V  = -\rho_\Lambda
\end{equation}
This means that the stress tensor can be thought of as made up of two components
-- one behaving like a pressure-less fluid, while the other having a negative
pressure. This suggests a possibility of providing a unified description of both dark matter
and dark energy using the same scalar field \cite{tptirth}. 
 
 When $\dot\phi$ is small (compared to $V$ in the case of quintessence or
   compared to unity in the case of tachyonic field), both these sources have $w\to -1$ and
   mimic a cosmological constant. When $\dot \phi \gg V$, the quintessence has $w\approx 1$ leading to
   $\rho_q\propto (1+z)^6$; the tachyonic field, on the other hand, has $w\approx 0$ for $\dot\phi\to 1$
   and behaves like non-relativistic matter. In both the cases, $-1<w<1$, though it is possible to construct more complicated scalar field Lagrangians  \cite{phantom} with even $w<-1$ describing
   what is called {\it phantom} matter.
  (For some  alternatives to scalar field models, based on brane world scenarios, see,
  for example, \cite{branes}.)
  
   Since  the quintessence field (or the tachyonic field)   has
   an undetermined free function $V(\phi)$, it is possible to choose this function
  in order to produce a given $H(a)$.
  To see this explicitly, let
   us assume that the universe has two forms of energy density with $\rho(a) =\rho_{\rm known}
  (a) + \rho_\phi(a)$ where $\rho_{\rm known}(a)$ arises from any known forms of source 
  (matter, radiation, ...) and
  $\rho_\phi(a) $ is due to a scalar field.  
  Let us first consider quintessence. Here,  the potential is given implicitly by the form
  \cite{ellis,tptachyon}
  \begin{equation}
  V(a) = \frac{1}{16\pi G} H (1-Q)\left[6H + 2aH' - \frac{aH Q'}{1-Q}\right]
  \label{voft}
   \end{equation} 
    \begin{equation}
    \phi (a) =  \left[ \frac{1}{8\pi G}\right]^{1/2} \int \frac{da}{a}
     \left[ aQ' - (1-Q)\frac{d \ln H^2}{d\ln a}\right]^{1/2}
    \label{phioft}
    \end{equation} 
   where $Q (a) \equiv [8\pi G \rho_{\rm known}(a) / 3H^2(a)]$ and prime denotes differentiation with respect to $a$.
   Given any
   $H(a),Q(a)$, these equations determine $V(a)$ and $\phi(a)$ and thus the potential $V(\phi)$.

   Every quintessence model studied in the literature can be obtained from these equations.
   We shall now briefly mention some examples:
   \begin{itemize}
  \item
  Power law expansion of the universe can be generated by
  a quintessence model with $V(\phi)=\phi^{-\alpha}$. In this case, the energy
  density of the scalar field varies as $\rho_\phi \propto t^{-2\alpha/(2+\alpha)} $;
  if the background density $\rho_{\rm bg}$ varies as $\rho_{\rm bg} \propto t^{-2}$,
  the ratio of the two energy densities changes as $(\rho_\phi/\rho_{\rm bg} =
  t^{4/(2+\alpha)}$).
  Obviously, the scalar field density can dominate over the background at late
  times for $\alpha >0$.  
  \item
  A different class of models arise if the potential is taken
  to be exponential with, say, $V(\phi) \propto \exp(-\lambda \phi/M_{\rm Pl})$.
  When $k=0$, both $\rho_\phi$ and $\rho_{\rm bg}$ scale in the same manner
  leading to 
  \begin{equation}
  \frac{\rho_\phi}{\rho_{\rm bg} +\rho_\phi} = \frac{3(1+w_{\rm bg})}{\lambda^2}
  \end{equation}
  where $w_{\rm bg}$  refers to the background parameter value. In this 
  case, the dark energy density is said to ``track'' the background energy
  density. While this could be a model for dark matter, there are 
  strong constraints on the total energy density of the universe
  at the epoch of nucleosynthesis. This requires $\Omega_\phi \lesssim 0.2$
  requiring dark energy to be sub dominant at all epochs. 
  
\item
  Many other forms of $H(a)$ can be reproduced by a combination of non-relativistic matter and a suitable form of scalar field with a potential $V(\phi)$. In fact, one can make the dark energy to vary with $a$ in an unspecified manner \cite{coop98} as $a^{-n}$. In this case we need $H^2(a)=H_0^2[ \Omega_{\rm NR} a^{-3}+(1-\Omega_{\rm NR})a^{-n}]$ which can arise if the universe is populated with non-relativistic matter with density parameter $\Omega_{\rm NR}$ and a scalar field with the potential,  determined using equations (\ref{voft}), (\ref{phioft}). We get 
  \begin{equation}
  V(\phi)=V_0 \sinh^{2n/(n-3)}[\alpha(\phi -\psi)]
   \end{equation}
where
  \begin{equation}
  V_0={(6-n)H_0^2\over 16\pi G}
  \left[ \frac{\Omega_{\rm NR}^n}{(1- \Omega_{\rm NR})^3}\right]^{\frac{1}{n-3}};\quad
  \alpha=(3-n)(2\pi G/n)^{1/2}
  \end{equation}
and $\psi$ is a constant.
  \end{itemize}
  
  Similar results exists for the tachyonic scalar field as well \cite{tptachyon}. For example, given
  any $H(a)$, one can construct a tachyonic potential $V(\phi)$ so that the scalar field is the 
  source for the cosmology. The equations determining $V(\phi)$  are now given by:
  \begin{equation}
  \phi(a) = \int \frac{da}{aH} \left(\frac{aQ'}{3(1-Q)}
   -{2\over 3}{a H'\over H}\right)^{1/2}
  \label{finalone}
  \end{equation}
   \begin{equation}
   V = {3H^2 \over 8\pi G}(1-Q) \left( 1 + {2\over 3}{a H'\over H}-\frac{aQ'}{3(1-Q)}\right)^{1/2}
   \label{finaltwo}
   \end{equation}
   Equations (\ref{finalone}) and (\ref{finaltwo}) completely solve the problem. Given any
   $H(a)$, these equations determine $V(a)$ and $\phi(a)$ and thus the potential $V(\phi)$. 
As an example, consider  a universe with power law expansion
   $a= t^n$. If it is populated only by  a tachyonic scalar field, then $Q=0$; further,
    $(a H'/H)$ in equation (\ref{finalone}) is a constant
     making $\dot \phi $  a constant. The complete solution
   is then given by
   \begin{equation}
   \phi(t) = \left({2\over 3n}\right)^{1/2} t + \phi_0; \quad
   V(t) = {3n^2\over 8\pi G}\left( 1- {2\over 3n}\right)^{1/2} {1\over t^2}
   \end{equation}
   where $n>(2/3)$.
   Combining the two, we find the potential to be
   \begin{equation}
    V(\phi) = {n\over 4\pi G}\left( 1- {2\over 3n}\right)^{1/2}
   (\phi - \phi_0)^{-2}
   \label{tachpot}
   \end{equation}
   For such a potential, it is possible to have arbitrarily rapid expansion with large $n$.   
   (For the cosmological model, based on this potential, see \cite{bjp}.)
    A wide variety of phenomenological models with time dependent
  \cc\ have been considered in the literature all of which can be 
   mapped to a 
  scalar field model with a suitable $V(\phi)$.

  While the scalar field models enjoy considerable popularity (one reason being they are easy to construct!)
  it is very doubtful whether they have helped us to understand the nature of the dark energy
  at any deeper level. These
  models, viewed objectively, suffer from several shortcomings:
  \begin{figure}[ht]
 \begin{center}
 \includegraphics[scale=0.5]{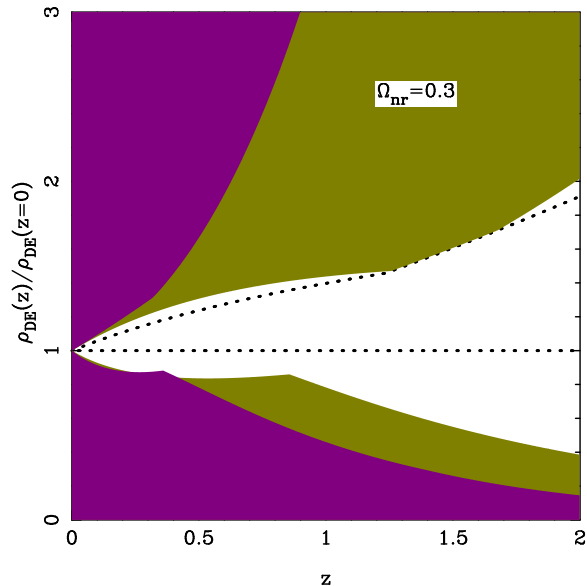}
 \end{center}
 \caption{Constraints on the possible variation of the dark energy density with redshift. The darker shaded region (magenta) is excluded by SN observations while the lighter shaded region (green) is excluded by WMAP observations. It is obvious that WMAP puts stronger constraints on the possible
 variations of dark energy density. The cosmological constant corresponds to the horizontal line
 at unity. The region between the dotted lines has $w>-1$ at all epochs. (For more details, see
 \cite{jbp}.) }
 \label{fig:bjp2ps}
 \end{figure}  
  \begin{itemize}
  \item
  They completely lack predictive power. As explicitly demonstrated above, virtually every form of $a(t)$ can be modeled by a suitable ``designer" $V(\phi)$.
  \item
  These models are  degenerate in another sense. The previous discussion  illustrates that even when $w(a)$ is known/specified, it is not possible to proceed further and determine
  the nature of the scalar field Lagrangian. The explicit examples given above show that there
  are {\em at least} two different forms of scalar field Lagrangians (corresponding to
  the quintessence or the tachyonic field) which could lead to
  the same $w(a)$. (See ref.\cite{tptirthsn1} for an explicit example of such a construction.)
  \item
  All the scalar field potentials require fine tuning of the parameters in order to be viable. This is obvious in the quintessence models in which adding a constant to the potential is the same as invoking a \cc. So to make the quintessence models work, \textit{we first need to assume the \cc\ is zero.} These models, therefore, merely push the cosmological constant problem to another level, making it somebody else's problem!.
  \item
  By and large, the potentials  used in the literature have no natural field theoretical justification. All of them are non-renormalisable in the conventional sense and have to be interpreted as a low energy effective potential in an adhoc manner.
  \item
  One key difference between \cc\ and scalar field models is that the latter lead to a $w(a)$ which varies with time. If observations have demanded this, or even if observations have ruled out $w=-1$ at the present epoch,
  then one would have been forced to take alternative models seriously. However, all available observations are consistent with \cc\ ($w=-1$) and --- in fact --- the possible variation of $w$ is strongly constrained \cite{jbp} as shown in Figure \ref{fig:bjp2ps}.
(Also see \cite{wconstraint}).
 \end{itemize}

 Given this situation, we shall now take a more serious look at the \cc\ as the source of dark energy in the universe.
 
 \section{...For the Snark was a Boojam, you see }
 
 If we assume that the dark energy in the universe is due to a \cc\, then we are introducing a second length scale, $L_\Lambda=H_\Lambda^{-1}$, into the theory  (in addition to the Planck length $L_P$) such that 
$ (L_P/L_\Lambda)\approx
 10^{-60}$. Such a universe will be asymptotically deSitter with $a(t)\propto \exp (t/L_\Lambda) $ at late times.
  We will now explore several peculiar features of such a universe.
  
  \begin{figure}
  \begin{center}
  \includegraphics[angle=-90,scale=0.6]{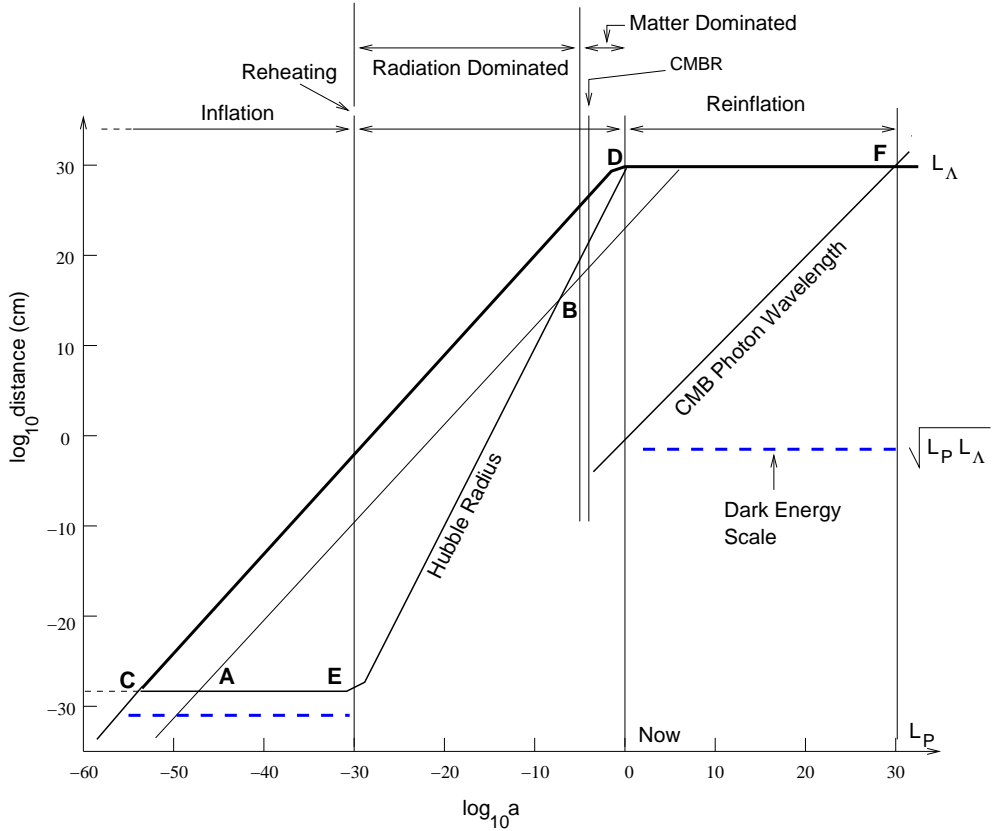}
  \end{center}
\caption{The geometrical structure of a universe with two length scales $L_P$ and $L_\Lambda$ corresponding to the Planck length and the cosmological constant \cite{plumian,bjorken}. Such a universe spends most of its time in two De Sitter phases which are (approximately) time translation invariant. The first De Sitter phase corresponds to the inflation and the second corresponds to the accelerated expansion arising from the cosmological constant. Most of the perturbations generated during the inflation will leave the Hubble radius (at some A, say) and re-enter (at B). However, perturbations which exit the Hubble radius
earlier than C will never re-enter the Hubble radius, thereby introducing a specific  dynamic range CE during the inflationary phase. The epoch F is characterized by the redshifted CMB temperature becoming equal to the De Sitter temperature $(H_\Lambda / 2\pi)$ which introduces another dynamic range DF in the accelerated expansion after which the universe is dominated by vacuum noise
of the De Sitter spacetime.}
\label{fig:tpplumian}
  \end{figure}
 
 Figure \ref{fig:tpplumian} summarizes these features \cite{plumian,bjorken}. Using the characteristic length scale of expansion,
 the Hubble radius $d_H\equiv (\dot a/a)^{-1}$, we can distinguish between three different phases of such a universe. The first phase is when the universe went through a inflationary expansion with $d_H=$ constant; the second phase is the radiation/matter dominated phase in which most of the standard cosmology operates and $d_H$ increases monotonically; the third phase is that of re-inflation (or accelerated expansion) governed by the cosmological constant in which $d_H$ is again a constant. The first and last phases are time translation invariant;
 that is, $t\to t+$ constant is an (approximate) invariance for the universe in these two phases. The universe satisfies the perfect cosmological principle and is in steady state during these phases!
 
 In fact, one can easily imagine a scenario in which the two deSitter phases (first and last) are of arbitrarily long duration \cite{plumian}. If  $\Omega_\Lambda\approx 0.7, \Omega_{DM}\approx 0.3$ the final deSitter phase \textit{does} last forever; as regards the inflationary phase, nothing prevents it from lasting for arbitrarily long duration. Viewed from this perspective, the in between phase --- in which most of the `interesting' cosmological phenomena occur ---  is  of negligible measure in the span of time. It merely connects two steady state phases of the universe.
 (In a way, this scenario provides the ultimate generalisation of the Copernican principle. It was
 well known that we are not in a special position in {\it space} in our universe. The composition of the universe also shows that we are not made of the most dominant constituent of the universe. Finally, in this picture, we are not even existing at a generic moment of {\it time} in the evolution of the universe!) 
 
    Given the two length scales $L_P$ and $L_\Lambda$, one can construct two energy scales
 $\rho_P=1/L_P^4$ and $\rho_\Lambda=1/L_\Lambda^4$ in natural units ($c=\hbar=1$). The first is, of course, the Planck energy density while the second one also has a natural interpretation. The universe which is asymptotically deSitter has a horizon and associated thermodynamics \cite{ghds} with a  temperature
 $T=H_\Lambda/2\pi$ and the corresponding thermal energy density $\rho_{thermal}\propto T^4\propto 1/L_\Lambda^4=
 \rho_\Lambda$. Thus $L_P$ determines the \textit{highest} possible energy density in the universe while $L_\Lambda$
 determines the {\it lowest} possible energy density in this universe. As the energy density of normal matter drops below this value, the thermal ambience of the deSitter phase will remain constant and provide the irreducible `vacuum noise'. Note that the dark energy density is the the geometric mean $\rho_{DE}=\sqrt{\rho_\Lambda\rho_P}$ between the two energy densities. If we define a dark energy length scale $L_{DE}$  such that $\rho_{DE}=1/L_{DE}^4$ then $L_{DE}=\sqrt{L_PL_\Lambda}$ is the geometric mean of the two length scales in the universe. The figure \ref{fig:tpplumian} also shows the variation of $L_{DE}$ by broken horizontal lines. 
 
 While the two deSitter phases can last forever in principle, there is a natural cut off length scale in both of them
 which makes the region of physical relevance to be finite \cite{plumian}. Let us first discuss the case of re-inflation in the late universe. 
 As the universe grows exponentially in the phase 3, the wavelength of CMBR photons are being redshifted rapidly. When the temperature of the CMBR radiation drops below the deSitter temperature (which happens when the wavelength of the typical CMBR photon is stretched to the $L_\Lambda$.)
 the universe will be essentially dominated by the vacuum thermal noise of the deSitter phase.
 This happens at the point marked F when the expansion factor is $a=a_F$ determined by the
  equation $T_0 (a_0/a_{F}) = (1/2\pi L_\Lambda)$. Let $a=a_\Lambda$ be the epoch at which
  cosmological constant started dominating over matter, so that $(a_\Lambda/a_0)^3=
  (\Omega_{DM}/\Omega_\Lambda)$. Then we find that the dynamic range of 
 DF is 
 \begin{equation}
\frac{a_F}{a_\Lambda} = 2\pi T_0 L_\Lambda \left( \frac{\Omega_\Lambda}{\Omega_{DM}}\right)^{1/3}
\approx 3\times 10^{30}
\end{equation}

 Interestingly enough, one can also impose a similar bound on the physically relevant duration of inflation. 
 We know that the quantum fluctuations generated during this inflationary phase could act as seeds of structure formation in the universe \cite{genofpert}. Consider a perturbation at some given wavelength scale which is stretched with the expansion of the universe as $\lambda\propto a(t)$.
 (See the line marked AB in Figure \ref{fig:tpplumian}.)
 During the inflationary phase, the Hubble radius remains constant while the wavelength increases, so that the perturbation will `exit' the Hubble radius at some time (the point A in Figure \ref{fig:tpplumian}). In the radiation dominated phase, the Hubble radius $d_H\propto t\propto a^2$ grows faster than the wavelength $ \lambda\propto a(t)$. Hence, normally, the perturbation will `re-enter' the Hubble radius at some time (the point B in Figure \ref{fig:tpplumian}).
 If there was no re-inflation, this will make {\it all} wavelengths re-enter the Hubble radius sooner or later.
 But if the universe undergoes re-inflation, then the Hubble radius `flattens out' at late times and some of the perturbations will {\it never} reenter the Hubble radius ! The limiting perturbation which just `grazes' the Hubble radius as the universe enters the re-inflationary phase is shown by the line marked CD in Figure \ref{fig:tpplumian}. If we use the criterion that we need the perturbation to reenter the Hubble radius, we get a natural bound on the duration of inflation which is of direct astrophysical relevance. This portion of the inflationary regime is marked by CE
 and can be calculated as follows: Consider  a perturbation which leaves the Hubble radius ($H_{in}^{-1}$) during the inflationary epoch at $a= a_i$. It will grow to the size $H_{in}^{-1}(a/a_i)$ at a later epoch. 
 We want to determine $a_i$ such that this length scale grows to 
   $L_\Lambda$ just when the dark energy starts dominating over matter; that is at
 the epoch $a=a_\Lambda = a_0(\Omega_{DM}/\Omega_{\Lambda})^{1/3}$. 
  This gives 
  $H_{in}^{-1}(a_\Lambda/a_i)=L_\Lambda$ so that $a_i=(H_{in}^{-1}/L_\Lambda)(\Omega_{DM}/\Omega_{\Lambda})^{1/3}a_0$. On the other hand, the inflation ends at 
  $a=a_{end}$ where $a_{end}/a_0 = T_0/T_{\rm reheat}$ where $T_{\rm reheat} $ is the temperature to which the universe has been reheated at the end of inflation. Using these two results we can determine the dynamic range of CE to be 
  \begin{equation}
\frac{a_{\rm end} }{a_i} = \left( \frac{T_0 L_\Lambda}{T_{\rm reheat} H_{in}^{-1}}\right)
\left( \frac{\Omega_\Lambda}{\Omega_{DM}}\right)^{1/3}=\frac{(a_F/a_\Lambda)}{2\pi T_{\rm reheat} H_{in}^{-1}} \cong 10^{25}
\end{equation} 
where we have used the fact that, for a GUTs scale inflation with $E_{GUT}=10^{14} GeV,T_{\rm reheat}=E_{GUT},\rho_{in}=E_{GUT}^4$
we have $2\pi H^{-1}_{in}T_{\rm reheat}=(3\pi/2)^{1/2}(E_P/E_{GUT})\approx 10^5$.
For a Planck scale inflation with $2\pi H_{in}^{-1} T_{\rm reheat} = \mathcal{O} (1)$, the phases CE and DF are approximately equal. The region in the quadrilateral CEDF is the most relevant part of standard cosmology, though the evolution of the universe can extend to arbitrarily large stretches in both directions in time. This figure is definitely telling us something regarding the time translation invariance of the universe (`the perfect cosmological principle') and --- more importantly ---
\textit{about the  breaking of this symmetry}, but it is not easy to translate this concept into a workable theory.

Let us now turn our attention to few of the many attempts to understand the \cc. This is, of course, a non-representative sample (dictated by personal bias!) and a host of other approaches exist in literature, some of which can be found in \cite{catchall}.

\subsection{Dark energy from a nonlinear correction term}

One of the \textit{least} esoteric ideas regarding the dark energy
is that the cosmological constant term in the FRW equations arises because we have not calculated the energy density driving the expansion of the universe correctly. The motivation for such a suggestion arises from the following fact:  The energy momentum tensor of the real universe, $T_{ab}(t,{\bf x})$ is inhomogeneous and anisotropic and will lead to a very complex metric $g_{ab}$ if only we could solve the exact Einstein's equations
$G_{ab}[g]=\kappa T_{ab}$.
The metric describing the large scale structure of the universe should be obtained by averaging this exact solution over a large enough scale, to get $\langle g_{ab}\rangle $. But what we actually do is to average the stress tensor {\it first} to get $\langle T_{ab}\rangle $ and {\it then} solve Einstein's equations. But since $G_{ab}[g]$ is  nonlinear function of the metric, $\langle G_{ab}[g]\rangle \neq G_{ab}[\langle g\rangle ]$ and there is a discrepancy. This is most easily seen by writing
\begin{equation}
G_{ab}[\langle g\rangle ]=\kappa [\langle T_{ab}\rangle  + \kappa^{-1}(G_{ab}[\langle g\rangle ]-\langle G_{ab}[g]\rangle )]\equiv \kappa [\langle T_{ab}\rangle  + T_{ab}^{corr}]
\end{equation}
If --- based on observations --- we take the $\langle g_{ab}\rangle $ to be the standard Friedman metric, this equation shows that it has, as its  source,  \textit{two} terms:
The first is the standard average stress tensor and the second is a purely geometrical correction term
$T_{ab}^{corr}=\kappa^{-1}(G_{ab}[\langle g\rangle ]-\langle G_{ab}[g]\rangle )$ which arises because of nonlinearities in the Einstein's theory that  leads to $\langle G_{ab}[g]\rangle \neq G_{ab}[\langle g\rangle ]$. If this term can mimic the \cc\ at large scales there will be no need for dark energy! Unfortunately, it is not easy to settle this question to complete satisfaction \cite{avgg}. One possibility is to use some analytic approximations to nonlinear perturbations (usually called non-linear scaling relations, see e.g. \cite{nsr}) to estimate this term.
This does not lead to a stress tensor that mimics dark energy (Padmanabhan, unpublished) but this is not a conclusive proof either way. We mention this mainly because this issue deserves more attention than it has received.

\subsection{Unimodular gravity}

Another possible way of addressing this issue is to simply eliminate from the gravitational theory those modes which couple to cosmological constant. If, for example, we have a theory in which the source of gravity is
$(\rho +p)$ rather than $(\rho +3p)$ in Eq. (\ref{nextnine}), then \cc\ will not couple to gravity at all. (The non linear coupling of matter with gravity has several subtleties; see eg. \cite{gravitonmyth}.) Unfortunately
it is not possible to develop a covariant theory of gravity using $(\rho +p)$ as the source. But we can achieve the same objective in different manner. Any metric $g_{ab}$ can be expressed in the form $g_{ab}=f^2(x)q_{ab}$ such that
${\rm det}\, q=1$ so that ${\rm det}\, g=f^4$. From the action functional for gravity
\begin{equation}
A=\frac{1}{16\pi G}\int d^4x (R -2\Lambda)\sqrt{-g}
=\frac{1}{16\pi G}\int d^4x R \sqrt{-g}-\frac{\Lambda}{8\pi G}\int d^4x f^4(x)
\end{equation}
it is obvious that the \cc\ couples {\it only} to the conformal factor $f$. So if we consider a theory of gravity in which $f^4=\sqrt{-g}$ is kept constant and only $q_{ab}$ is varied, then such a model will be oblivious of
direct coupling to \cc. If the action (without the $\Lambda$ term) is varied, keeping ${\rm det}\, g=-1$, say, then one is lead to a {\it unimodular theory of gravity} with the equations of motion 
$R_{ab}-(1/4)g_{ab}R=\kappa(T_{ab}-(1/4)g_{ab}T)$ with zero trace on both sides. Using the Bianchi identity, it is now easy to show that this is equivalent to a theory with an {\it  arbitrary} \cc. That is, \cc\ arises as an undetermined integration constant in this model \cite{unimod}. 

While this is interesting, we need an extra physical principle to fix its value.
One possible way of doing this is to  interpret the $\Lambda$ term in the action as a Lagrange multiplier for the proper volume of the spacetime. Then it is reasonable to choose the \cc\ such that the total proper volume of the universe is equal to a specified number. While this will lead to a \cc\ which has the correct order of magnitude, it has several obvious problems. First, the proper four volume of the universe is infinite unless we make the spatial sections compact and restrict the range of time integration. Second, this will lead to a dark energy density  which varies as $t^{-2}$ (corresponding to $w= -1/3$ ) which is ruled out by observations.

\subsection{Scale dependence of the vacuum energy}

The conventional discussion of the relation between cosmological constant and vacuum energy density is based on
evaluating the zero point energy of quantum fields with an ultraviolet cutoff and using the result as a 
source of gravity.
Any reasonable cutoff will lead to a vacuum energy density $\rho_{\rm vac}$ which is unacceptably high. 
This argument,
however, is too simplistic since the zero point energy --- obtained by summing over the
$(1/2)\hbar \omega_k$ --- has no observable consequence in any other phenomena and can be subtracted out by redefining the Hamiltonian. The observed non trivial features of the vacuum state of QED, for example, arise from the {\it fluctuations} (or modifications) of this vacuum energy rather than the vacuum energy itself. 
This was, in fact,  known fairly early in the history of cosmological constant problem and, in fact, is stressed by Zeldovich \cite{zeldo} who explicitly calculated one possible contribution to {\it fluctuations} after subtracting away the mean value.
This
suggests that we should consider   the fluctuations in the vacuum energy density in addressing the 
cosmological constant problem. 

If the vacuum probed by the gravity can readjust to take away the bulk energy density $\rho_P\simeq L_P^{-4}$, quantum \textit{fluctuations} can generate
the observed value $\rho_{\rm DE}$. One of the simplest models \cite{tpcqglamda} which achieves this uses the fact that, in the semiclassical limit, the wave function describing the universe of proper four-volume ${\cal V}$ will vary as
$\Psi\propto \exp(-iA_0) \propto 
 \exp[ -i(\Lambda_{\rm eff}\mathcal V/ L_P^2)]$. If we treat 
  $(\Lambda/L_P^2,{\cal V})$ as conjugate variables then uncertainty principle suggests $\Delta\Lambda\approx L_P^2/\Delta{\cal V}$. If
the four volume is built out of Planck scale substructures, giving $ {\cal V}=NL_P^4$, then the Poisson fluctuations will lead to $\Delta{\cal V}\approx \sqrt{\cal V} L_P^2$ giving
    $ \Delta\Lambda=L_P^2/ \Delta{\mathcal V}\approx1/\sqrt{{\mathcal V}}\approx   H_0^2
 $. (This idea can be a more quantitative; see \cite{tpcqglamda}).

Similar viewpoint arises, more formally, when we study the question of \emph{detecting} the energy
density using gravitational field as a probe.
 Recall that an Unruh-DeWitt detector with a local coupling $L_I=M(\tau)\phi[x(\tau)]$ to the {\it field} $\phi$
actually responds to $\langle 0|\phi(x)\phi(y)|0\rangle$ rather than to the field itself \cite{probe}. Similarly, one can use the gravitational field as a natural ``detector" of energy momentum tensor $T_{ab}$ with the standard coupling $L=\kappa h_{ab}T^{ab}$. Such a model was analysed in detail in ref.~\cite{tptptmunu} and it was shown that the gravitational field responds to the two point function $\langle 0|T_{ab}(x)T_{cd}(y)|0\rangle $. In fact, it is essentially this fluctuations in the energy density which is computed in the inflationary models \cite{inflation} as the seed {\it source} for gravitational field, as stressed in
ref.~\cite{tplp}. All these suggest treating the energy fluctuations as the physical quantity ``detected" by gravity, when
one needs to incorporate quantum effects.  
If the \cc\ arises due to the energy density of the vacuum, then one needs to understand the structure of the quantum vacuum at cosmological scales. Quantum theory, especially the paradigm of renormalization group has taught us that the energy density --- and even the concept of the vacuum
state --- depends on the scale at which it is probed. The vacuum state which we use to study the
lattice vibrations in a solid, say, is not the same as vacuum state of the QED. Using this feature, it is possible to construct systems in condensed matter physics \cite{volovikilya} wherein the quantity analogous to
vacuum energy density has to vanish on the average because of dynamical reasons.

 In fact, it seems \textit{inevitable} that in a universe with two length scale $L_\Lambda,L_P$, the vacuum
 fluctuations will contribute an energy density of the correct order of magnitude $\rho_{DE}=\sqrt{\rho_\Lambda\rho_P}$. The hierarchy of energy scales in such a universe has \cite{plumian,tpvacfluc}
 the pattern
 \begin{equation}
\rho_{\rm vac}={\frac{1}{ L^4_P}}    
+{\frac{1}{L_P^4}\left(\frac{L_P}{L_\Lambda}\right)^2}  
+{\frac{1}{L_P^4}\left(\frac{L_P}{L_\Lambda}\right)^4}  
+  \cdots 
\end{equation}  
 The first term is the bulk energy density which needs to be renormalised away (by a process which we  do not understand at present); the third term is just the thermal energy density of the deSitter vacuum state; what is interesting is that quantum fluctuations in the matter fields \textit{inevitably generate} the second term.

The key new ingredient arises from the fact that the properties of the vacuum state  depends on the scale at which it is probed and it is not appropriate to ask questions without specifying this scale. 
 (These ideas have been developed more generally in ref. \cite{holo}.)
 If the spacetime has a cosmological horizon which blocks information, the natural scale is provided by the size of the horizon,  $L_\Lambda$, and we should use observables defined within the accessible region. 
The operator $H(<L_\Lambda)$, corresponding to the total energy  inside
a region bounded by a cosmological horizon, will exhibit fluctuations  $\Delta E$ since vacuum state is not an eigenstate of 
{\it this} operator. The corresponding  fluctuations in the energy density, $\Delta\rho\propto (\Delta E)/L_\Lambda^3=f(L_P,L_\Lambda)$ will now depend on both the ultraviolet cutoff  $L_P$ as well as $L_\Lambda$.  
 To obtain
 $\Delta \rho_{\rm vac} \propto \Delta E/L_\Lambda^3$ which scales as $(L_P L_\Lambda)^{-2}$
 we need to have $(\Delta E)^2\propto L_P^{-4} L_\Lambda^2$; that is, the square of the energy fluctuations
 should scale as the surface area of the bounding surface which is provided by the  cosmic horizon.  
 Remarkably enough, a rigorous calculation \cite{tpvacfluc} of the dispersion in the energy shows that
 for $L_\Lambda \gg L_P$, the final result indeed has  the scaling 
 \begin{equation}
 (\Delta E )^2 = c_1 \frac{L_\Lambda^2}{L_P^4} 
 \label{deltae}
 \end{equation}
 where the constant $c_1$ depends on the manner in which ultra violet cutoff is imposed.
 Similar calculations have been done (with a completely different motivation, in the context of 
 entanglement entropy)
 by several people and it is known that the area scaling  found in Eq.~(\ref{deltae}), proportional to $
L_\Lambda^2$, is a generic feature \cite{area}.
For a simple exponential UV-cutoff, $c_1 = (1/30\pi^2)$ but  cannot be computed
 reliably without knowing the full theory.
  We thus find that the fluctuations in the energy density of the vacuum in a sphere of radius $L_\Lambda$ 
 is given by 
 \begin{equation}
 \Delta \rho_{\rm vac}  = \frac{\Delta E}{L_\Lambda^3} \propto L_P^{-2}L_\Lambda^{-2} \propto \frac{H_\Lambda^2}{G}
 \label{final}
 \end{equation}
 The numerical coefficient will depend on $c_1$ as well as the precise nature of infrared cutoff 
 radius (like whether it is $L_\Lambda$ or $L_\Lambda/2\pi$ etc.). It would be pretentious to cook up the factors
 to obtain the observed value for dark energy density. 
 But it is a fact of life that a fluctuation of magnitude $\Delta\rho_{vac}\simeq H_\Lambda^2/G$ will exist in the
energy density inside a sphere of radius $H_\Lambda^{-1}$ if Planck length is the UV cut off. {\it One cannot get away from it.}
On the other hand, observations suggest that there is a $\rho_{vac}$ of similar magnitude in the universe. It seems 
natural to identify the two, after subtracting out the mean value by hand. Our approach explains why there is a \textit{surviving} cosmological constant which satisfies 
$\rho_{DE}=\sqrt{\rho_\Lambda\rho_P}$
 which ---  in our opinion --- is {\it the} problem. 

There is a completely different way of interpreting this result based on some imaginative ideas suggested by Bjorken
\cite{bjorken} recently. The key idea  is to parametrise the universes by the value of $L_\Lambda$ which they have. It is a fixed, pure number for each universe in an ensemble of universes but all the other parameters of the physics are assumed to be  correlated with
$L_\Lambda$. This is motivated by a series of arguments in ref. \cite{bjorken} and, in this approach, $\rho_{vac}\propto L_\Lambda^{-2}$ almost by definition; the hard work was in determining how other parameters scale with $L_\Lambda$. In the approach suggested here, a dynamical interpretation of the scaling  $\rho_{vac}\propto L_\Lambda^{-2}$ is given as due to vacuum fluctuations of  fields. We now reinterpret each member of of the ensemble of universes as having zero energy density for vacuum (as any decent vacuum should have) but the effective $\rho_{vac}$ arises from the quantum fluctuations {\it with the correct scaling}. One can then invoke standard anthropic-like arguments (but with very significant differences as stressed in ref. \cite{bjorken} ) to choose a range for the size of our universe. This appears to be much more attractive way of interpreting the result.

Finally, to be fair, this attempt should be
 judged in the backdrop of other suggested solutions almost all of which require introducing
 extra degrees of freedom in the form of scalar fields, modifying gravity or introducing higher dimensions etc. {\it and} fine tuning the potentials.  At a fundamental level such approaches are unlikely to provide the final solution.

 \section*{Acknowledgement}
 
 I thank K. Subramanian for two decades of discussion about various aspects of cosmological constant and for sharing and reinforcing the view that any quick-fix solution to this problem will be futile. I also thank 
 Apoorva Patel for useful discussions.

\end{document}